\documentclass[twocolumn,prb]{revtex4}
\usepackage{amssymb,bbm,bm}
\usepackage{graphicx}
\usepackage{bm}
\usepackage{amsmath}
\usepackage[usenames]{color}
\usepackage{epsfig}
\usepackage{hyperref}
\usepackage{subfigure}
\usepackage[sans]{dsfont}
\hypersetup{colorlinks=true, citecolor=blue,
linkcolor=blue,urlcolor=blue }

\usepackage{amsmath,amsfonts,amssymb,amsthm,bm}
\usepackage{float}
\usepackage{hyperref}
\usepackage{graphicx}
\usepackage{subfigure}

\newcommand{\qab}{q_{\alpha \beta}}
\newcommand{\ma}{m_{\alpha}}
\newcommand{\db}{\mathrm{D}}
\newcommand{\df}{\mathrm{d}}

\begin{document}

\title{Magnetic proximity in a coupled Ferromagnet-Spin glass system}
\author{Fateme Izadi}
\author{Reza Sepehrinia}\email{sepehrinia@ut.ac.ir}
\affiliation{Department of Physics, University of Tehran, Tehran 14395-547, Iran}

\begin{abstract}
We study the competition between ferromagnetic and spin glass phases using a system of coupled infinite-range Ising and Sherrington-Kirkpatrick models. We obtain the replica-symmetric solution for the free energy of this system in terms of magnetization and Edwards-Anderson order parameters in both subsystems. Using these order parameters we are able to identify different phases in the system and determine which phase is dominant for different strengths of the coupling between two subsystems. We observe that both subsystems are in the same phase although with different order parameters. The phase boundary between paramagnetic-ferromagnetic and paramagnetic-spin glass phases is more or less similar to that in the Sherrington-Kirkpatrick phase diagram. But the boundary between ferromagnetic-spin glass phases becomes qualitatively different as the coupling between the two subsystems changes. Remarkably, for intermediate values of the coupling, this phase boundary is such that there could be a reentrant transition between spin glass and ferromagnetic phases by increasing the temperature. We found that for some range of the coupling, the second order transition between these phases turns into first order at a tricritical point. Further, we carry out the stability analysis by considering deviations from the replica symmetric solution.

\end{abstract}
\maketitle

\section{Introduction}

Combinations of materials with different magnetic properties are of tremendous utility in electronics and spintronics for magnetic recording and designing devices such as spin valves, pseudo-spin valves, magnetic tunnel junctions, magnetic sensors, etc. Various novel properties of these composite systems stem from the interface effects. Exchange bias and magnetic proximity are particularly important phenomena that are common in many of these systems \cite{manna2014two}. The exchange bias is a unidirectional magnetic anisotropy that first was observed in a system of a coupled ferromagnet (FM) and antiferromagnet (AFM) \cite{meiklejohn1956new} and by now it is observed in a variety of other systems \cite{kishimoto1979coercivity,berkowitz1988toward,cain1990investigation,van1995study,zheng2004exchange,binek2006exchange,huang2007coexistence}. It has been used, for instance, as a stabilizer in magnatic recording heads and spin valves. Magnetic proximity is a rather general phenomenon \cite{rader1994fe,yang1995polarization,tomaz1997induced,schwickert1998magnetic,van2000difference,yi2001microstructure,leighton2002coercivity,skumryev2003beating,grimsditch2003exchange,lenz2007magnetic,van2007proximity,maccherozzi2008evidence} which is capable of inducing and changing magnetic properties such as coercivity, blocking temperature, transition temperature to various ordered phases and etc which also has various applications.

Despite the technological impact, a theoretical understanding of these phenomena has posed many challenges for a long time  \cite{kiwi2001exchange,radu2008exchange,nogues1999exchange}. Earlier models of exchange bias failed to quantitatively predict the experimental observations \cite{meiklejohn1962exchange,neel1967ferro}. Over the years more sophisticated models were introduced to explain these phenomena \cite{malozemoff1987random,mauri1987simple,koon1997calculations,schulthess1998consequences}. One of the ingredients that seemed to play an important role was imperfections in the interface and impurities or defects in the bulk of AFM component \cite{schulthess1998consequences,miltenyi2000diluted,nowak2002modeling,spray2006exchange}. This led to the conjecture that the exchange bias would also occur in the FM layer coupled to a spin glass (SG). This was confirmed in several experiments and Monte Carlo simulations \cite{schlenker1986magnetic,westerholt2003exchange,wang2004surface,usadel2009exchange}. Later experimental work revealed an interesting feature in the FM/SG bilayer that the sign of the exchange bias changes depending on the thickness of the SG layer \cite{ali2007exchange}. Even though many experimental and theoretical works have been devoted to the exchange bias effect it is still a subject of intense research.

A theoretical study of the magnetic proximity has also been done for several systems \cite{manna2014two} however much less attention was given to this phenomenon compared to the exchange bias effect. Particularly the proximity effect in a coupled FM and SG system is interesting according to ongoing experiments and the above mentioned peculiar properties of this system. Motivated by recent experimental works \cite{ali2007exchange,chi2019spin,yu2020temperature,rui2015cooling,chi2016exchange,chi2020role} we have done a theoretical study on the equilibrium phase diagram of a coupled FM and SG system. The main question that we try to answer is to what extent the SG order can penetrate into the FM and vice versa.

We use a minimal model consisting of an infinite-range Ising FM and Sherrington-Kirkpatrick (SK) spin glass \cite{sherrington1975solvable,mezard1987spin,nishimori2001statistical,dotsenko2005introduction}, exchange-coupled to each other. The infinite range interactions that we consider are equivalent to the mean-field approximation for a real finite-dimensional material. Excluding the fluctuations in  mean-field approximation causes the overestimation of the critical temperature and also incorrect critical exponents, however one expects to find a qualitatively correct picture of the phase diagram. Also the effect of the thickness of the materials in contact can not be studied using this simplified model, rather it presumably mimics the equilibrium phase near the interface.
We utilize the standard replica method to obtain the ensemble-averaged free energy and carry out stability analysis and obtain the de Almeida-Thouless (AT) line of the model for different strengths of the coupling between two subsystems.
\section{Model}
As we mentioned, we would like to consider an Ising model coupled with an SK model. But in order to compute the spin glass order parameter in the Ising model, first we consider a system of two coupled SK models and after deriving the state equations we set the width of the coupling distribution on one of them to be zero. So we start with the Hamiltonian
\begin{equation}\label{eq1}
H = - \sum_{i < j} J_{ij} \sigma_i \sigma_j -  \sum_{i<j} J^{\prime}_{ij} \tau_i \tau_j -D \sum_{i} \sigma_i \tau_i,
\end{equation}
where $ \sigma_i=\pm 1 $ and $ \tau_i= \pm 1 $  are Ising spins defined on the first and the second model, respectively, and the quenched random couplings $ J_{ij} $ are independent and identically distributed with a Gaussian distribution
\begin{equation}\label{eq2}
P(J_{ij}) = \frac{1}{J} \sqrt{\frac{N}{2 \pi}} \exp \left(- \frac{N}{2J^2} \Big(J_{ij} - \frac{J_0}{N} \Big)^2 \right).
\end{equation}
Similarly, $ {J}^{\prime}_{ij} $ is drawn from a Gaussian distribution function $ P^{\prime}(J^{\prime}_{ij}) $ with mean $ J_I $ and variance $ J' $. We use the replica method to calculate the ensemble-averaged free energy per spin which is based on the following well-known identity 	
\begin{equation}\label{eq3}
-\beta f= \lim_{N \to \infty} \lim_{n \to 0} \frac{1}{N n} \ln \overline{Z^n},
\end{equation}
where $ Z= \mathrm{Tr} \ e^{-\beta H} $ is the partition function as a function of inverse temperature $ \beta = 1/kT $. Here we use the overbar for the average over the quenched random couplings and $\langle \cdot \rangle$ for thermal averages. As we mentioned, in the end we will set $ J'=0 $ so we will have an SK model coupled to an infinite-range ferromagnetic Ising model. Following the method used for the SK model \cite{sherrington1975solvable,mezard1987spin,nishimori2001statistical,dotsenko2005introduction}, we find the free energy in the limit $ N \to \infty $ as follows
\begin{widetext}
 \begin{eqnarray}\label{eq4}
 -\beta f & = & \lim_{n \to 0} \Bigg\{-\frac{\beta^2 J^2 }{2n}  \sum_{\alpha < \beta}  {\qab^s}^2 - \frac{\beta J_0 }{2n} \sum_{\alpha} {\ma^s}^2 -\frac{\beta^2 J'^2 }{2n}  \sum_{\alpha < \beta} {\qab^i}^2 - \frac{\beta J_I }{2n} \sum_{\alpha} {\ma^i}^2  + \frac{\beta^2 (J^2 + J'^2) }{4} \nonumber\\
 & &+ \frac{1}{n} \log \mathrm{Tr} \exp \Bigg( \beta^2 J^2 \sum_{\alpha < \beta} \qab^s   \sigma_{\alpha}  \sigma_{\beta}  + \beta J_0 \sum_{\alpha}  \ma^s  \sigma_{\alpha} + \beta^2 J'^2 \sum_{\alpha < \beta} \qab^i  \tau_{\alpha}  \tau_{\beta} + \beta J_I \sum_{\alpha}  \ma^i \tau_{\alpha}  + \beta D  \sum_{\alpha} \sigma_{\alpha} \tau_{\alpha} \Bigg) \Bigg\}. \nonumber \\
\end{eqnarray}
\end{widetext}
The indices $ \alpha $ and $ \beta $ run from 1 to n are the replica indexes. The variables $ \ma^s $,$ \qab^s $ and $ \ma^i $,$ \qab^i $ turn out to represent the ferromagnetic and the spin glass order parameter for the two subsystems
\begin{eqnarray}\label{eq6}
 m_s &=& \langle \sigma_{\alpha} \rangle,  \qquad  q_s = \langle \sigma_{\alpha} \sigma_{\beta} \rangle \notag, \\
 m_i &=& \langle \tau_{\alpha} \rangle,  \qquad q_i = \langle \tau_{\alpha} \tau_{\beta} \rangle.
\end{eqnarray}

\subsection{Replica symmetric (RS) solution }
Assuming the replica symmetric ansatz, $ \qab = q $, $ \ma = m $, and by taking the limit $ n \to 0 $, the free energy in Eq. (\ref{eq4}) becomes
\begin{widetext}
\begin{eqnarray}\label{eq5}
 - \beta f &=& \frac{1}{4} \beta^2 J^2 (1-q_s)^2  - \frac{1}{2}  \beta J_0 m_s^2 + \frac{1}{4} \beta^2 J'^2 (1-q_i)^2  - \frac{1}{2} \beta J_I m_i^2  \nonumber \\
 & &  + \int \db z_1 \int \db z_2 \log \Big[ e^{\beta D} 2 \cosh(\beta J \sqrt{q_s} z_1 + \beta J_0 m_s + \beta J' \sqrt{q_i} z_2 + \beta J_I m_i)  \nonumber \\
 & & \hspace{6.5cm} + e^{- \beta D} 2 \cosh(\beta J \sqrt{q_s} z_1 + \beta J_0 m_s - \beta J' \sqrt{q_i} z_2 - \beta J_I m_i) \Big],
\end{eqnarray}
\end{widetext}
where $  \db z \equiv \df z \exp (-z^2 / 2) / \sqrt{2 \pi} $ is the Gaussian measure. The extremization of the free energy with respect to $ m_s $, $ q_s $, $ m_i $ and $ q_i $ gives the equations of state as follows
\begin{equation}\label{eq7}
m_s = \int \db z \  \frac{B_{+}}{A}, \qquad m_i = \int \db z \  \frac{B_{-}}{A},
\end{equation}
\begin{equation}\label{eq8}
q_s = \int \db z \  \Big(\frac{B_{+}}{A}\Big)^2, \qquad q_i = \int \db z \  \Big(\frac{B_{-}}{A}\Big)^2,
\end{equation}
where
\begin{eqnarray*}
A &=& e^{\beta D}  \cosh(\beta ( J \sqrt{q_s} z + J_0 m_s + J_I m_i)) \\
& &+ e^{- \beta D} \cosh(\beta ( J \sqrt{q_s} z + J_0 m_s - J_I m_i)),
\end{eqnarray*}
and
\begin{eqnarray*}
B_{\pm} &=& e^{\beta D} \sinh(\beta ( J \sqrt{q_s} z + J_0 m_s + J_I m_i)) \\
& &\pm e^{-\beta D} \sinh(\beta ( J \sqrt{q_s} z + J_0 m_s - J_I m_i)).
\end{eqnarray*}
We have set $J' = 0$ in the last step.
\subsection{Limiting cases}
In order to see the consistency of the results, we consider the limiting cases. For $ D=0 $  the subsystems become decoupled and the equations of the standard Ising and SK models are obtained as expected
\begin{equation}\label{}
m_i = \tanh(\beta J_I m_i), \ \ q_i =m_i^2
\end{equation}
\begin{eqnarray}\label{}
m_s &=& \int \db z \  \tanh(\beta J \sqrt{q_s} z + \beta  J_0 m_s ), \\
q_s &=& \int \db z \  \tanh^2(\beta J \sqrt{q_s} z + \beta  J_0 m_s ).
\end{eqnarray}
On the other limiting case, $ D \to \infty $, the large coupling constant makes the spins $\sigma_i$ and $\tau_i$  align. The equations of state reduce to
\begin{eqnarray}
m &=& \int \db z  \tanh(\beta J \sqrt{q_s} z + \beta ( J_0 + J_I ) m ),  \\
q &=& \int \db z  \tanh^2 \left(\beta  J \sqrt{q_s} z + \beta ( J_0 + J_I ) m  \right),
\end{eqnarray}
where $m=m_i=m_s$ and $q=q_i=q_s$. These are the equations of state for an SK model with a shifted mean coupling constant $ J_0 +‌J_I $. Therefore, for the strong coupling, the Hamiltonian Eq. (\ref{eq1}) reduces to that of the SK model with a shifted mean coupling constant.
\subsection{Stability of the RS solution}
It is well known that the replica symmetric solution might be unstable \cite{de1978stability}. To test the validity of the RS assumption we use the method developed in Ref. \cite{de1978stability} which is based on expanding the free energy to second order in deviations from the replica symmetric solution. This quadratic form must be positive definite which leads to the stability condition (see Appendix):
\begin{equation}\label{eq9}
\left(\frac{kT}{J}\right)^2 > \int \db z \left( 1- \left( \frac{B_{+}}{A}\right)^2 \right)^2.
\end{equation}
This inequality determines the region in the parameter space in which the RS solution is stable. The boundary between the stable and unstable region is called the AT line which we will determine below.
Again at the limit $ D \to \infty $ this equation reduces to the stability condition of the SK model \cite{sherrington1975solvable}.
\section{PHASE DIAGRAMS}\label{sec3}
\begin{figure}[t]
\begin{center}
\subfigure[]{
\includegraphics[scale=0.7]{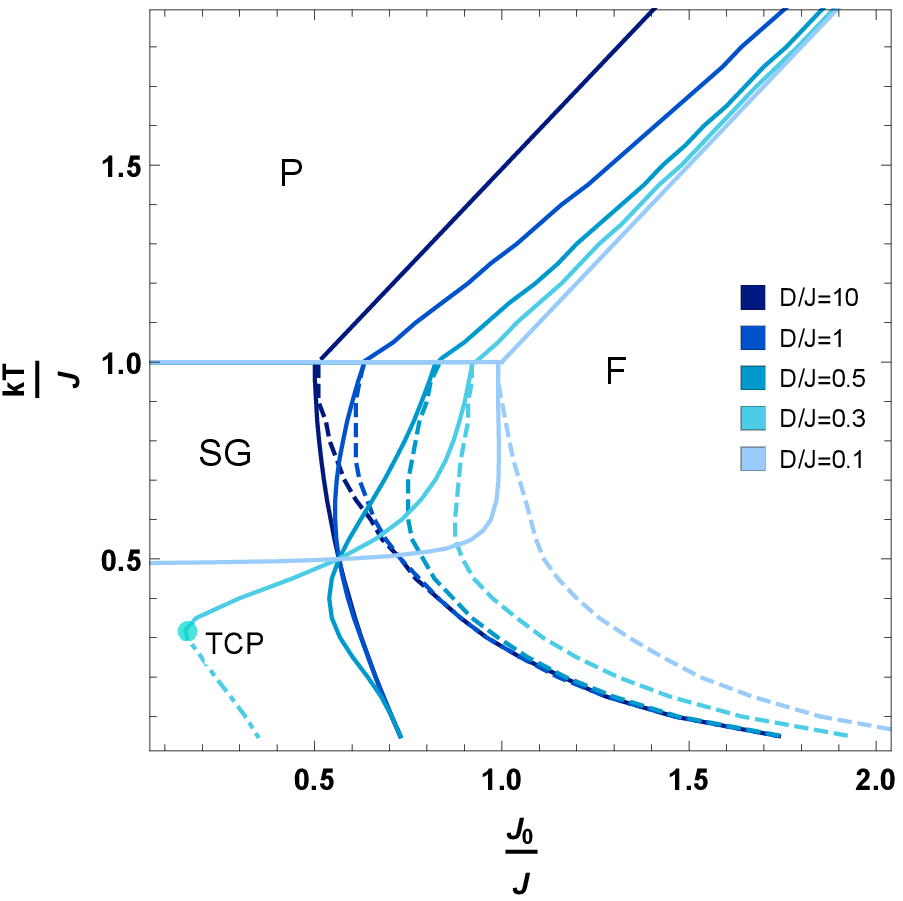}
\label{kT-J0-1}
}
\hspace*{0.1cm}
\subfigure[]{
\includegraphics[scale=0.7]{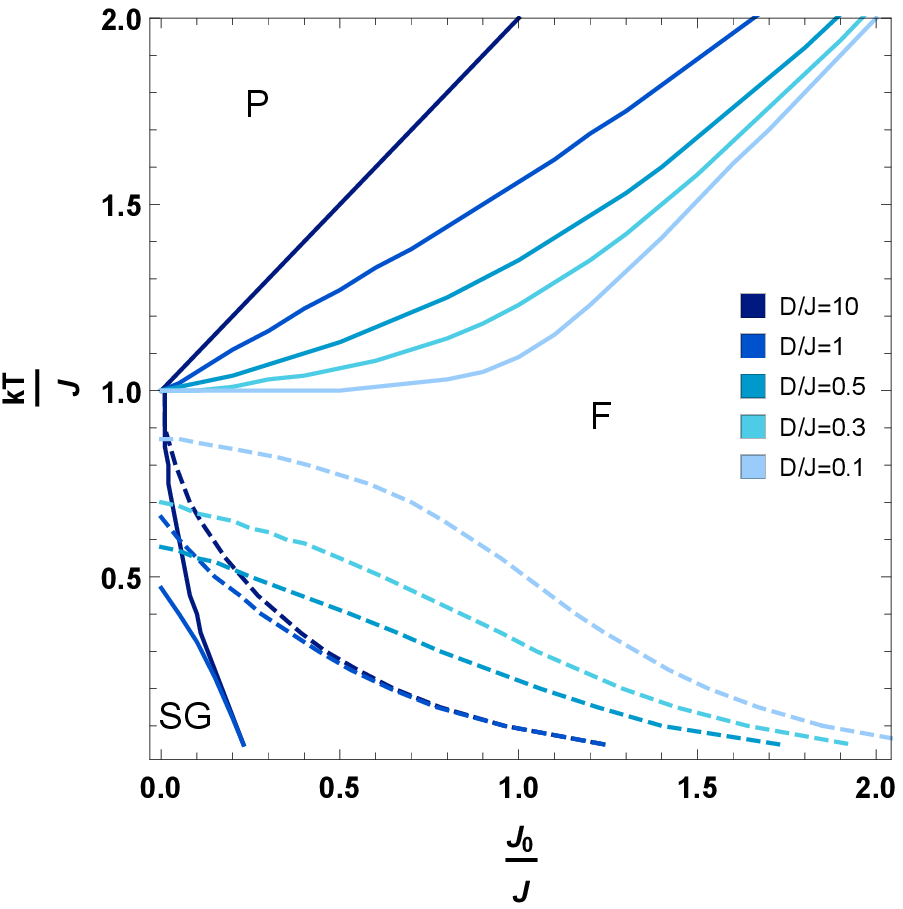}
\label{kT-J0-2}
}
\end{center}
\caption{The $ (J_0/J, kT/ J) $ phase diagram for several values of $D$ and with
\subref{kT-J0-1}   $ J_I/J = 0.5 $ and
\subref{kT-J0-2}  $ J_I/J = 1 $ exhibits a paramagnetic (P), a ferromagnetic (F), and a spin-glass (SG) phase. The dashed lines are the AT line under which the RS solution is unstable. The dashed-dotted line in \subref{kT-J0-1} indicates the first order transition which is separated from the second order transition line by a tricritical point (TCP).
\label{kT-J0}
}
\end{figure}
The equilibrium state of the system can be determined with four dimensionless parameters $ kT/J, D/J, J_0/J, J_I/J $.

The first observation from the equations of state is that two subsystems are in the same phase although with different values of order parameters. To see this, first we consider $m_i=0$ then from Eq. (\ref{eq7}) we have $m_i=\tanh (\beta D)\int Dz \tanh(\beta ( J \sqrt{q_s} z + J_0 m_s ))=0$. This integral will be zero only if $m_s=0$. Conversely, if we consider $m_s=0$ then Eq. (\ref{eq7}) becomes $m_s=\int Dz \tanh(\beta ( J \sqrt{q_s} z + K ))=0$, where $\tanh \beta K=\tanh (\beta J_I m_i) \tanh(\beta D) $. Similarly the integral is zero if $K=0$ and the equation for $K$ requires $m_i=0$. Therefore vanishing the magnetization in one of the subsystems implies the vanishing of the magnetization in the other. In other words, both subsystems will be in the ferromagnetic phase ($\mathbf{F}$) or not at the same time. Now we assume $m_i=m_s=0$ in which case the Eq. (\ref{eq8}) gives $q_s=\int Dz \tanh^2(\beta  J \sqrt{q_s} z)$ and $q_i=\tanh^2(\beta D)\int Dz \tanh^2(\beta  J \sqrt{q_s} z)$. As we can see, in the non-$\mathbf{F}$ phase, the SG order parameter in the SK model is independent of the coupling $D$ and in the Ising model $q_i=\tanh^2(\beta D) q_s$. Again the SG order parameter is zero or nonzero in both models at the same time. If both $q$'s are nonzero we have the spin-glass phase ($\mathbf{SG}$) phase and if they are zero we have the paramagnetic phase ($\mathbf{P}$).

At sufficiently high temperatures the system will be in the $\mathbf{P}$ phase with $ m_s=m_i=q_s=q_i=0 $. At low temperatures, there is a competition between the $\mathbf{F}$ phase with $ m_i \neq 0 , m_s \neq 0 , q_i \neq 0 , q_s \neq 0  $ , and $\mathbf{SG}$ phase with $ m_i = m_s = 0 , q_i \neq 0 , q_s \neq 0  $.

\textit{$\mathbf{P}$-$\mathbf{F}$ phase boundary.} According to the above definition of the phases as the $\mathbf{P}$-$\mathbf{F}$ phase boundary is approached from the $\mathbf{F}$ phase, all the order parameters vanish. Setting the $q_s$ equal to zero in Eq. (\ref{eq7})and expanding the right hand side for small $m_s$ and $m_i$ we obtain
\begin{eqnarray}\label{}
m_s&=&\beta J_0 m_s+\beta\tanh(\beta D) J_I m_i, \\
m_i&=&\beta\tanh(\beta D) J_0 m_s+\beta J_I m_i.
\end{eqnarray}
To have a nonzero solution for $m_s$ and $m_i$, the determinant of the matrix of the coefficients must vanish which gives the following relation
\begin{equation}\label{PF}
(\beta J_0-1)(\beta J_I-1)-\beta^2 \tanh^2(\beta D) J_I J_0=0.
\end{equation}
The $\mathbf{P}$-$\mathbf{F}$ phase boundary changes from the line $kT = J_0$ for $D=0$ to $kT = J_I + J_0$ for $D \rightarrow \infty$ (see Fig. \ref{kT-J0}).

\textit{$\mathbf{P}$-$\mathbf{SG}$ phase boundary.} In both $\mathbf{P}$ and $\mathbf{SG}$ phases magnetization vanishes. So we set $m_s=m_i=0$ in Eq. (\ref{eq8}) and expand the right hand side for small $q$ which gives
\begin{equation}\label{PSG}
\beta^2 J^2=1.
\end{equation}
This phase boundary is independent of $D$.

\textit{$\mathbf{F}$-$\mathbf{SG}$ phase boundary.} To obtain the $\mathbf{F}$-$\mathbf{SG}$ phase boundary we solve Eqs. \eqref{eq7}, and \eqref{eq8} numerically. It turns out that, in some cases, the magnetization changes discontinuously along a part of this boundary, i.e., the transition is first order.

AT \textit{line}. We obtain the region of stability of the RS solution using the relation Eq. \eqref{eq9}. Below the AT line, the replica symmetric solution is unstable.
\subsection{Asymmetric distribution}
Figure \ref{kT-J0} depicts the phase diagram of the model in $ (J_0/J, kT/ J) $ plane for different values of coupling $ D $. The dashed lines indicate the AT lines under which the RS solution is unstable. As it was shown, the phase diagram approaches the SK model phase diagram, as coupling $D$ increases. At small and intermediate couplings $D$, the shape of the $\mathbf{SG}$-$\mathbf{F}$ phase boundary is different considerably from the SK model.
\begin{figure}[h]
\begin{center}
\includegraphics[scale=0.4]{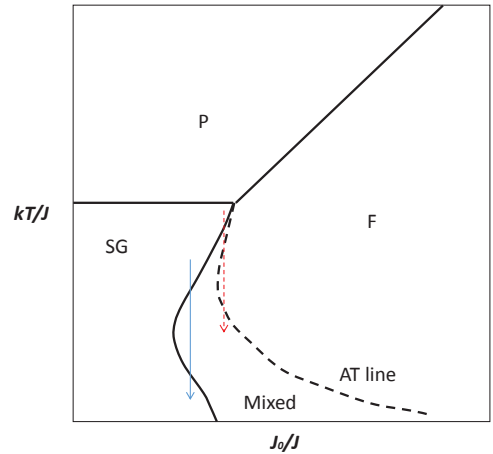}
\end{center}
\caption{Phase diagram for $D/J=0.5$. Solid blue arrow shows the reentrant transition $\mathbf{SG}$-$\mathbf{F}$-$\mathbf{SG}$ by decreasing temperature. Dashed red arrow shows the transition $\mathbf{SG}$-$\mathbf{F}$-$\mathbf{Mixed}$}
\label{reentrant}
\end{figure}
\begin{figure}[b]
\begin{center}
\includegraphics[scale=0.8]{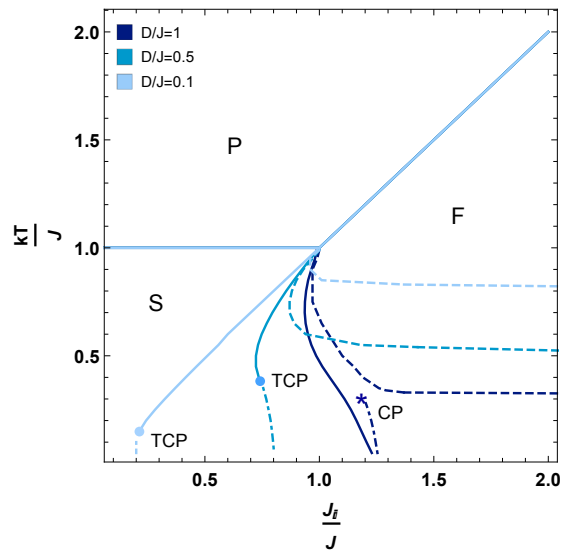}
\end{center}
\caption{The $ (J_I/J, kT/ J) $ phase diagram for several values of $D/J$ and $ J_0 = 0 $. The dashed lines are the AT lines. The dash-dotted lines indicate the first order transition which are separated from the second order transition line by a tricritical point (TCP) and terminate at the critical point (CP).}
\label{kT-Ji}
\end{figure}
\begin{figure}[t]
\begin{center}
\includegraphics[scale=0.8]{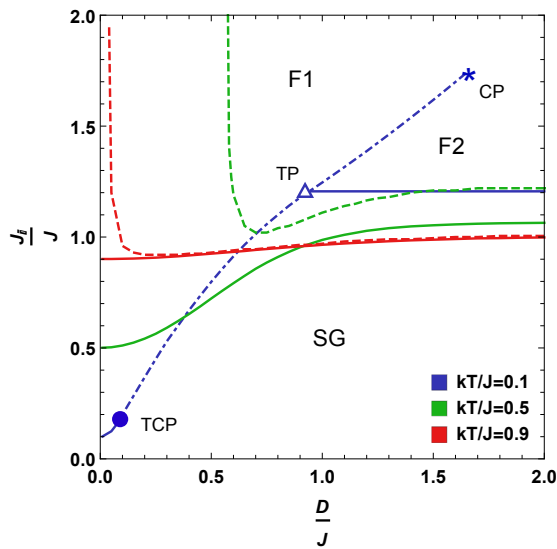}
\end{center}
\caption{The $ (D/J, J_I/ J) $ phase diagram for several values of $kT/J$ and $ J_0 = 0 $. The dashed lines are the AT lines. The dash-dotted line indicates the first order transition which start at a tricritical point (TCP) and terminate at the critical point (CP). Inside the ferromagnetic phase, magnetization changes discontinuously across the first order transition line where we have distinguished phases on two sides by $\mathbf{F_1}$ and $\mathbf{F_2}$. The first order transition line intersects the continuous transition line at a triple point (TP) which is indicated by the symbol $\triangle$. }
\label{Ji-D}
\end{figure}

\textit{Reentrant transition}.
A remarkable feature of the $\mathbf{SG}$-$\mathbf{F}$ boundary for intermediate coupling e.g. $D/J=0.5$ in Fig. \ref{kT-J0-1}, is that there could be a transition from the $\mathbf{SG}$ phase to $\mathbf{F}$ phase and back into the $\mathbf{SG}$ phase by decreasing the temperature (Fig. \ref{reentrant}, blue solid arrow), which is called the reentrant transition. This is in contrast to the reentrant transition in the original SK model which is known as the transition from the paramagnetic phase to a ferromagnetic one and then the transition from a ferromagnetic to the spin-glass phase (see Figs. 29 and 44 in Ref. \cite{binder1986spin}).

Under replica symmetry breaking (RSB), the $\mathbf{SG}$-$\mathbf{F}$ boundary in the original SK model, becomes vertical  (see Fig. 1 in Ref. \cite{toulouse1980mean}, Fig. 49 in Ref. \cite{binder1986spin}, and Fig. 8.4 in Ref. \cite{nishimori2001statistical}). Therefore the reentrant transition in the original SK model is the artifact of the replica symmetry ansatz. In our model, the replica symmetry breaking may result in a similar situation, however,
as we can see in Fig. \ref{reentrant}, the stable ferromagnetic region, which is shown by the AT line, is such that there could be a transition from $\mathbf{SG}$ to $\mathbf{F}$ and then from the $\mathbf{F}$ phase to another phase below the AT line (see Ref. \cite{fischer1993spin} Chap. 11). This has been shown by the dashed red arrow in Fig. \ref{reentrant}. The latter phase which has features of both $\mathbf{SG}$ and $\mathbf{F}$ phases is called the Mixed phase \cite{toulouse1980mean}.

If the Ising interaction $ J_I/J $ is strong enough, the $\mathbf{SG}$ phase may completely disappear (Fig. \ref{kT-J0-2}).

\subsection{Symmetric distribution}
In order to see the competition between $\mathbf{F}$ and $\mathbf{SG}$ phases more clearly, we reduce the number of parameters by assuming the symmetric distribution ($J_0=0$). For $J_0=0$ and $D=0$, the SK model must be in the $\mathbf{SG}$ phase if $kT<J$ and make a transition to the $\mathbf{P}$ phase by increasing the temperature. The Ising model must be in the $\mathbf{F}$ phase if $kT<J_I$ and in the $\mathbf{P}$ phase if $kT>J_I$.
\begin{figure*}[t]
\begin{center}
\subfigure[]{
\includegraphics[scale=0.5]{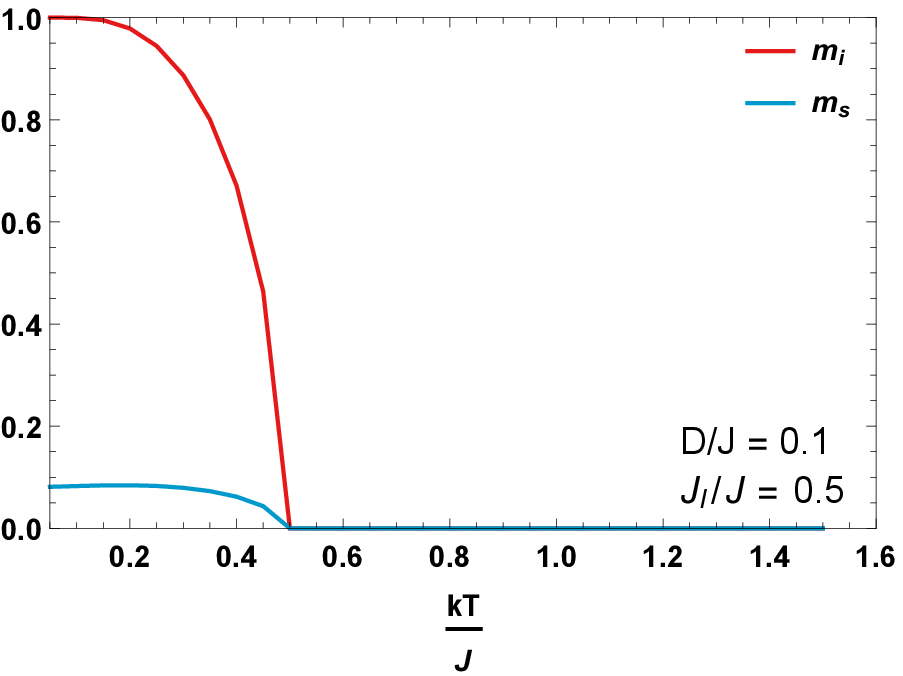}
\label{mq-1}
}
\subfigure[]{
\includegraphics[scale=0.5]{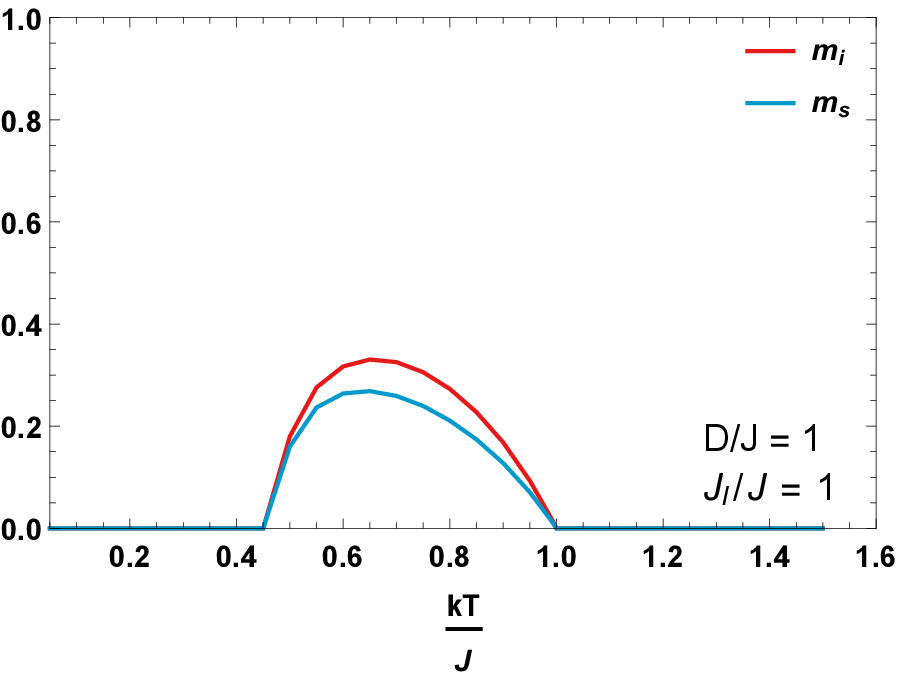}
\label{mq-2}
}
\subfigure[]{
\includegraphics[scale=0.5]{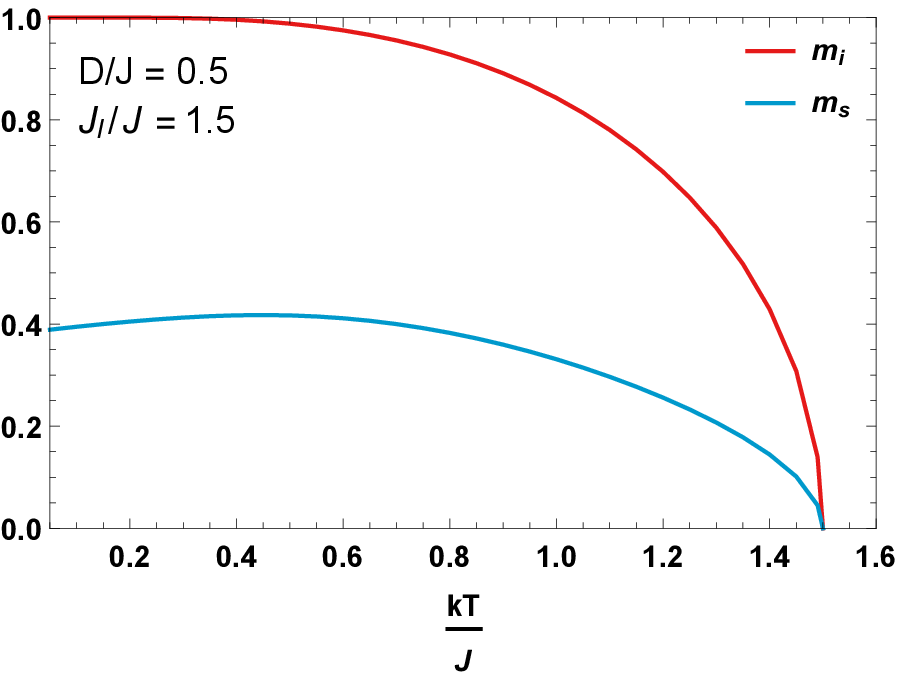}
\label{mq-3}
}
\subfigure[]{
\includegraphics[scale=0.5]{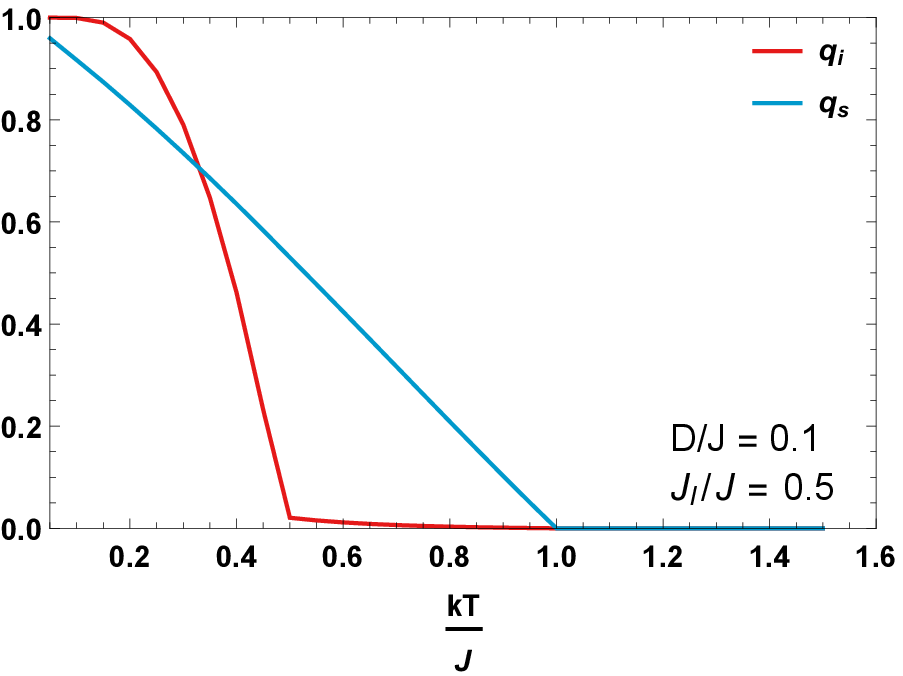}
\label{mq-4}
}
\subfigure[]{
\includegraphics[scale=0.5]{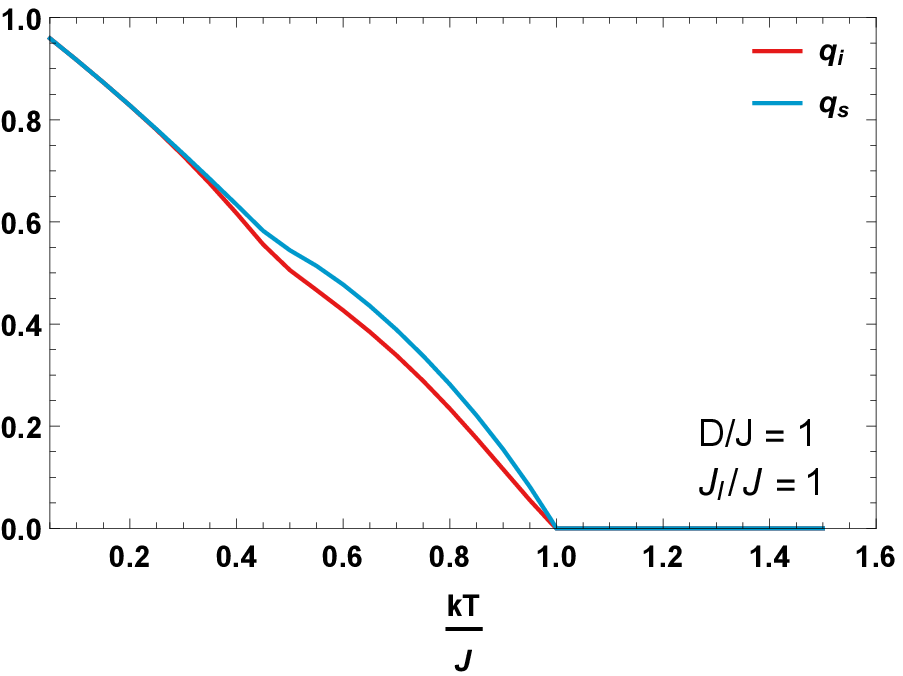}
\label{mq-5}
}
\subfigure[]{
\includegraphics[scale=0.5]{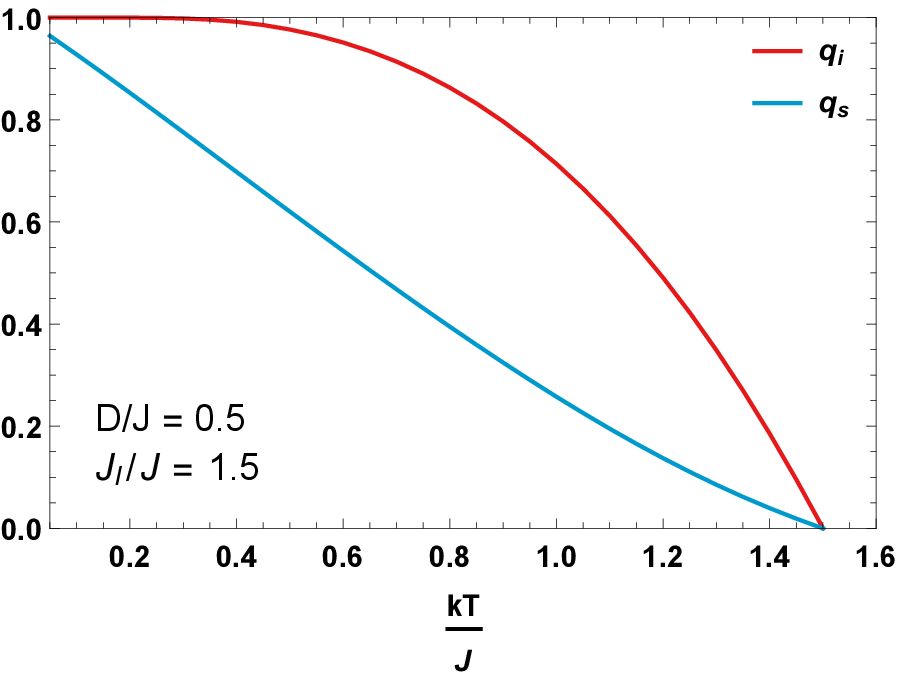}
\label{mq-6}
}
\end{center}
\caption{Order parameters of the model  magnetization \subref{mq-1},\subref{mq-2},\subref{mq-3} and spin-glass order parameter \subref{mq-4},\subref{mq-5},\subref{mq-6} respectively with
$ J_I/J = 0.5 $ and $ D/J = 0.1 $
, $ J_I/J = 1 $ and $ D/J = 1 $
and $ J_I/J = 1 $ and $ D/J = 1 $.
\label{mq}
}
\end{figure*}
Now for a nonzero coupling $D$, the phase diagram in Fig. \ref{kT-Ji} shows that which phase is dominant. From Eqs. (\ref{PF}), and (\ref{PSG}), for $J_0=0$, the $\mathbf{P}$-$\mathbf{F}$ phase boundary is given by $\beta J_I=1$ and $\mathbf{P}$-$\mathbf{SG}$ phase boundary will be $\beta J=1$, which are independent of $D$ in agreement with the phase diagram obtained by a numerical solution of the equations of state. For small values of $J_I$ the $\mathbf{SG}$ phase and for large values of $J_I$ the $\mathbf{F}$ phase is dominant. This is an expected result but the temperature dependence of the critical value of $J_I$ is nontrivial and as we mentioned causes the reentrant transition. As the coupling $D$ increases the phase diagram becomes the phase diagram of the SK model with shifted mean coupling equal to $J_I$.

The other way of representing the phase diagram is in terms of parameters $ (D/J,J_I/J) $ for different temperatures. Figure \ref{Ji-D} shows the low-temperature phase diagram. As it can be seen in this phase diagram, for any given coupling $D$, there is a critical value of $J_I$ under which the system makes a transition from the $\mathbf{F}$ to $\mathbf{SG}$ phase. One property that can be observed is the temperature dependence of this critical value. For the weak coupling $D$, the critical value of $J_I$ increases by increasing the temperature, and at strong coupling the trend is reversed.
For weak coupling and low temperature, the system is in the $\mathbf{F}$ phase even for small Ising coupling, but at strong coupling, the $\mathbf{SG}$ phase is dominant for relatively higher values of $J_I$.

The dashed lines in Fig. \ref{Ji-D} are the AT lines. The dashed-dotted line indicates the first order transition which starts at a tricritical point (TCP) and terminates at the critical point (CP). Inside the ferromagnetic phase, magnetization changes discontinuously across the first order transition line where we have distinguished phases on two sides by $\mathbf{F_1}$ and $\mathbf{F_2}$ where both are ferromagnetic but with different magnetizations. The first order transition line intersects the continuous transition line at a triple point (TP).

The other noticeable aspect is that regardless of how small the coupling $D$ is, both subsystems will be in the same phase. It is instructive to compare the values of the order parameters in two subsystems, especially for weak coupling. The order parameters of the system are shown in Fig. \ref{mq} for $ J_0 = 0 $ and different values of $D/J$ and $J_I/J$. For a weak coupling $D/J=0.1$, $D/J_I=0.2$ (Figs. \ref{mq-1}, and \ref{mq-4}), at low temperature the system is in the $\mathbf{F}$ phase, however, the $m_s$ is small compared to $m_i$. By increasing the temperature a transition to the $\mathbf{SG}$ phase occurs. In this phase $q_i$ is small compared to $q_s$. This shows the weak permeability of the ferromagnetic order and spin-glass order, respectively, between subsystems when the coupling is small. For $D/J=1$, $J_I/J=1$ (Figs. \ref{mq-2}, and \ref{mq-5}), the low temperature phase is $\mathbf{SG}$, by increasing the temperature transition to the $\mathbf{F}$ phase and again back to the $\mathbf{SG}$ phase occurs. This is a reentrant transition which we mentioned before. As we can see, the order parameters are of the same order in the two subsystems. The next plot is for $D/J=0.5$, $J_I/J=1.5$ (Figs. \ref{mq-3}, and \ref{mq-6}). In this case, due to the larger value of $J_I$, the $\mathbf{SG}$ phase has disappeared and the order parameters in the $\mathbf{F}$ phase are of the same order.

\section{CONCLUSIONS}\label{sec4}

We performed a detailed study of the phase diagram of the coupled infinite-range Ising and SK models using the replica method. We obtained the replica symmetric solutions for the magnetization and spin glass order parameter and investigated the stability of these solutions using the de Almeida-Thouless approach. The whole system exhibits three phases and in all cases, both subsystems are in the same phase for nonzero coupling $D$,
although with different values of the order parameters.

The $\mathbf{P}$-$\mathbf{SG}$ phase boundary in general and the $\mathbf{P}$-$\mathbf{F}$ phase boundary in the symmetric case ($J_0=0$) are independent of coupling $D$. The shape of the $\mathbf{F}$-$\mathbf{SG}$ phase boundary changes by varying the coupling $D$ in a nontrivial way, such that there could be a reentrant transition from $\mathbf{SG}$ to $\mathbf{F}$ and back into the $\mathbf{SG}$ phase by decreasing temperature (see Fig. \ref{reentrant}), in contrast to the reentrant $\mathbf{P}$-$\mathbf{F}$-$\mathbf{SG}$ transition in the original SK model. The transition between the $\mathbf{F}$ and $\mathbf{SG}$ phases becomes discontinuous, i.e. first order, for some range of parameters. The first order transition line extends into the $\mathbf{F}$ phase where it separates two ferromagnetic phases $\mathbf{F_1}$ and $\mathbf{F_2}$ with different magnetizations and terminates in a critical point.

For strong Ising coupling $J_I$ the $\mathbf{SG}$ phase and for sufficiently weak Ising coupling the $\mathbf{F}$ phase may disappear entirely.

For a weak coupling ($D\ll J,J_I$), in the $\mathbf{F}$ phase a small magnetization is induced in the SK model, and in the $\mathbf{SG}$ phase a small spin-glass order is induced in the Ising model, while for the strong coupling the order parameters in the two subsystems are of the same order.

From AT analysis we find that in the present model, similar to the SK model, the $\mathbf{SG}$-$\mathbf{F}$ phase boundary is in the unstable region. This is shown in Figs. \ref{kT-J0}, \ref{kT-Ji}, and \ref{Ji-D} where the $\mathbf{SG}$-$\mathbf{F}$ boundaries (solid) are located on the left of the corresponding AT line (dashed), where the replica symmetric solution is unstable. However, the stable region is such that a similar reentrant transition is more likely to survive under the replica symmetry breaking, namely the transition from the $\mathbf{SG}$ to $\mathbf{F}$ phase and then to the mixed phase (see Fig. \ref{reentrant}).

The model that we have introduced here is simple enough to carry out detailed calculations and provides the insight to understand the magnetic behavior of the FM/SG bilayer.
\section{acknowledgments}
We would like to acknowledge financial support from the research council of University of Tehran. This work is based upon research funded by Iran National Science Foundation (INSF) under project No.4005950.
\section{APPENDIX}
	
In this appendix, we derive the stability condition of the replica symmetric solution. By rescaling the variables as
\begin{eqnarray*}
& \beta J \qab^s = y^{\alpha \beta} \qquad \sqrt{\beta J_0} \ma^s = x^{\alpha } \\
& \beta J' \qab^i = v^{\alpha \beta} \qquad \sqrt{\beta J_I} \ma^i = w^{\alpha}
\end{eqnarray*}
one can rewrite the Eq. \eqref{eq4}
\begin{widetext}
\begin{eqnarray}\label{eq10}
f &=& -\frac{\beta (J^2 + J'^2) }{4} - \lim_{n \to 0} \frac{1}{\beta n} \Bigg\{-\frac{1}{2} \sum_{\alpha < \beta}  (y^{\alpha \beta})^2 - \frac{1}{2} \sum_{\alpha} (x^{\alpha})^2  -\frac{1}{2} \sum_{\alpha < \beta}  (v^{\alpha \beta})^2
-\frac{1}{2} \sum_{\alpha} (w^{\alpha})^2  \nonumber \\
& &+ \log \mathrm{Tr} \exp \Bigg( \beta J \sum_{\alpha < \beta} y^{\alpha \beta}   \sigma_{\alpha}  \sigma_{\beta}  + \sqrt{\beta J_0} \sum_{\alpha}  x^{\alpha}  \sigma_{\alpha}
+ \beta J' \sum_{\alpha < \beta} v^{\alpha \beta}  \tau_{\alpha}  \tau_{\beta}  + \sqrt{\beta J_I } \sum_{\alpha}  w^{\alpha}  \tau_{\alpha}  + \beta D  \sum_{\alpha} \sigma_{\alpha} \tau_{\alpha} \Bigg) \Bigg\}. \nonumber \\
\end{eqnarray}
\end{widetext}
To examine the stability, one must expand the free energy in small deviations around the RS solution
\begin{eqnarray*}
&y^{\alpha \beta} =y + \eta^{\alpha \beta} \qquad  x^{\alpha} = x + \varepsilon^{\alpha } \\
&v^{\alpha \beta}= v + \rho^{\alpha \beta} \qquad  w^{\alpha} = w + \nu^{\alpha }
\end{eqnarray*}
where $ x , y , v $ and $ w $ are the RS solution given in Eq. \eqref{eq7} and Eq. \eqref{eq8}. So the  second-order term of Eq. \eqref{eq10} with respect to $ \eta , \varepsilon , \rho $ and $ \nu $ is
\begin{widetext}
\begin{eqnarray}
\delta f &=& \frac12 \sum_{\alpha ,\beta} \Big\{ \delta_{\alpha \beta} - \beta J_0 \Big( \langle \sigma_{\alpha} \sigma_{\beta} \rangle - \langle \sigma_{\alpha} \rangle \langle \sigma_{\beta} \rangle \Big) \Big\} \varepsilon^{\alpha} \varepsilon^{\beta}+
\frac12 \sum_{\alpha < \beta} \sum_{\gamma < \delta} \Big\{ \delta_{(\alpha \beta)(\gamma \delta)} - \beta^2 J^2 ( \langle \sigma_{\alpha} \sigma_{\beta} \sigma_{\gamma} \sigma_{\delta}\rangle -
\langle \sigma_{\alpha} \sigma_{\beta} \rangle \langle \sigma_{\gamma} \sigma_{\delta} \rangle) \Big\} \eta^{\alpha \beta} \eta^{\gamma \delta} \nonumber \\
& &+\frac12 \sum_{\alpha ,\beta} \Big\{ \delta_{\alpha \beta} - \beta J_I \Big( \langle \tau_{\alpha} \tau_{\beta} \rangle - \langle \tau_{\alpha} \rangle \langle \tau_{\beta} \rangle \Big) \Big\} \nu^{\alpha} \nu^{\beta} +
\frac12 \sum_{\alpha < \beta} \sum_{\gamma < \delta} \Big\{ \delta_{(\alpha \beta)(\gamma \delta)} - \beta^2 J'^2 ( \langle \tau_{\alpha} \tau_{\beta} \tau_{\gamma} \tau_{\delta} \rangle -
\langle \tau_{\alpha} \tau_{\beta} \rangle \langle \tau_{\gamma} \tau_{\delta} \rangle) \Big\} \rho^{\alpha \beta} \rho^{\gamma \delta}  \nonumber \\
& & +\beta J \sqrt{\beta J_0} \sum_{\delta} \sum_{\alpha < \beta} ( \langle \sigma_{\alpha} \sigma_{\beta} \rangle \langle \sigma_{\delta} \rangle - \langle \sigma_{\alpha} \sigma_{\beta} \sigma_{\delta} \rangle )  \eta^{\alpha \beta} \varepsilon^{\delta} +
\beta J \sqrt{\beta J_I} \sum_{\delta} \sum_{\alpha < \beta} ( \langle \sigma_{\alpha} \sigma_{\beta} \rangle \langle \tau_{\delta} \rangle - \langle \sigma_{\alpha} \sigma_{\beta} \tau_{\delta} \rangle )  \eta^{\alpha \beta} \nu^{\delta}  \nonumber \\
& & +\beta J' \sqrt{\beta J_0} \sum_{\delta} \sum_{\alpha < \beta} ( \langle \tau_{\alpha} \tau_{\beta} \rangle \langle \sigma_{\delta} \rangle - \langle \tau_{\alpha} \tau_{\beta} \sigma_{\delta} \rangle )  \rho^{\alpha \beta} \varepsilon^{\delta} +
\beta J' \sqrt{\beta J_I} \sum_{\delta} \sum_{\alpha < \beta} ( \langle \tau_{\alpha} \tau_{\beta} \rangle \langle \tau_{\delta} \rangle - \langle \tau_{\alpha} \tau_{\beta} \tau_{\delta} \rangle )  \rho^{\alpha \beta} \nu^{\delta}  \nonumber \\
& & +\beta \sqrt{J_I J_0} \sum_{\alpha , \beta} ( \langle \tau_{\alpha} \rangle \langle \sigma_{\beta} \rangle - \langle \tau_{\alpha} \sigma_{\beta} \rangle )  \nu^{\alpha} \varepsilon^{\beta} +
\beta^2 J J' \sum_{\alpha < \beta} \sum_{\gamma < \delta}  (  \langle \tau_{\alpha} \tau_{\beta} \rangle \langle \sigma_{\gamma} \sigma_{\delta} \rangle - \langle \tau_{\alpha} \tau_{\beta} \sigma_{\gamma} \sigma_{\delta}\rangle ) \rho^{\alpha \beta} \eta^{\gamma \delta} \nonumber
\end{eqnarray}
\end{widetext}
The matrix of coefficients of this quadratic form, $ G $ , must have positive eigenvalues for the stable solution. This matrix has different kinds of the matrix elements. The second order terms in $ \varepsilon $ have the form
\begin{eqnarray*}
G_{\alpha \alpha} &=& 1 - \beta J_0 ( 1 - \langle \sigma_{\alpha} \rangle^2 ) \equiv A \\
G_{\alpha \beta} &=&  - \beta J_0 (  \langle \sigma_{\alpha} \sigma_{\beta} \rangle - \langle \sigma_{\alpha} \rangle^2 ) \equiv B.
\end{eqnarray*}
The second order terms in $ \eta $ are
\begin{eqnarray*}
G_{(\alpha \beta)(\alpha \beta)} &=& 1 - \beta^2 J^2 ( 1 - \langle \sigma_{\alpha} \sigma_{\beta} \rangle^2 ) \equiv P \\
G_{(\alpha \beta)(\alpha \delta)} &=&  - \beta^2 J^2 ( \langle \sigma_{\beta} \sigma_{\delta} \rangle - \langle \sigma_{\alpha} \sigma_{\beta} \rangle^2 ) \equiv Q \\
G_{(\alpha \beta)(\delta \gamma)} &=&  - \beta^2 J^2 ( \langle \sigma_{\alpha} \sigma_{\beta} \sigma_{\delta} \sigma_{\gamma} \rangle - \langle \sigma_{\alpha} \sigma_{\beta} \rangle^2) \equiv R.
\end{eqnarray*}
similarly the second order terms in $ \nu $ are
\begin{eqnarray*}
G_{\alpha \alpha} &=& 1 - \beta J_I ( 1 - \langle \tau_{\alpha} \rangle^2 ) \equiv S \\
G_{\alpha \beta} &=&  - \beta J_I (  \langle \tau_{\alpha} \tau_{\beta} \rangle - \langle \tau_{\alpha} \rangle^2 ) \equiv T.
\end{eqnarray*}
and the second order terms in $ \rho $ are
\begin{eqnarray*}
G_{(\alpha \beta)(\alpha \beta)} &=& 1 - \beta^2 J'^2 ( 1 - \langle \tau_{\alpha} \tau_{\beta} \rangle^2 ) \equiv L \\
G_{(\alpha \beta)(\alpha \delta)} &=&  - \beta^2 J'^2 ( \langle \tau_{\beta} \tau_{\delta} \rangle - \langle \tau_{\alpha} \tau_{\beta} \rangle^2 ) \equiv M \\
G_{(\alpha \beta)(\delta \gamma)} &=&  - \beta^2 J'^2 ( \langle \tau_{\alpha} \tau_{\beta} \tau_{\delta} \tau_{\gamma} \rangle - \langle \tau_{\alpha} \tau_{\beta} \rangle^2) \equiv N.
\end{eqnarray*}
The cross terms in $ \varepsilon $ and $ \eta $ have the form
\begin{eqnarray*}
G_{\alpha (\alpha \beta)} &=& \beta J \sqrt{\beta J_0} ( \langle \sigma_{\alpha} \rangle \langle \sigma_{\alpha} \sigma_{\beta} \rangle -  \langle \sigma_{\beta} \rangle ) \equiv C \\
G_{\delta (\alpha \beta)} &=&  \beta J \sqrt{\beta J_0} ( \langle \sigma_{\alpha} \rangle \langle \sigma_{\alpha} \sigma_{\beta} \rangle - \langle \sigma_{\alpha} \sigma_{\beta} \sigma_{\delta} \rangle ) \equiv D.
\end{eqnarray*}
The cross terms in $ \nu $ and $ \eta $ are
\begin{eqnarray*}
G_{\alpha (\alpha \beta)} &=& \beta J \sqrt{\beta J_I} ( \langle \tau_{\alpha} \rangle \langle \sigma_{\alpha} \sigma_{\beta} \rangle -  \langle \sigma_{\alpha} \sigma_{\beta} \tau_{\alpha} \rangle ) \equiv E \\
G_{\delta (\alpha \beta)} &=&  \beta J \sqrt{\beta J_I} ( \langle \tau_{\alpha} \rangle \langle \sigma_{\alpha} \sigma_{\beta} \rangle - \langle \sigma_{\alpha} \sigma_{\beta} \tau_{\delta} \rangle ) \equiv F.
\end{eqnarray*}
The cross terms in $ \varepsilon $ and $ \rho $ are
\begin{eqnarray*}
G_{\alpha (\alpha \beta)} &=& \beta J' \sqrt{\beta J_0} ( \langle \sigma_{\alpha} \rangle \langle \tau_{\alpha} \tau_{\beta} \rangle -  \langle \tau_{\alpha} \tau_{\beta} \sigma_{\alpha} \rangle ) \equiv O \\
G_{\delta (\alpha \beta)} &=&  \beta J' \sqrt{\beta J_0} ( \langle \sigma_{\delta} \rangle \langle \tau_{\alpha} \tau_{\beta} \rangle - \langle \tau_{\alpha} \tau_{\beta} \sigma_{\delta} \rangle ) \equiv W.
\end{eqnarray*}
The cross terms in $ \nu $ and $ \rho $ are
\begin{eqnarray*}
G_{\alpha (\alpha \beta)} &=& \beta J' \sqrt{\beta J_I} ( \langle \tau_{\alpha} \rangle \langle \tau_{\alpha} \tau_{\beta} \rangle -  \langle \tau_{\beta} \rangle ) \equiv U \\
G_{\delta (\alpha \beta)} &=&  \beta J' \sqrt{\beta J_I} ( \langle \tau_{\delta} \rangle \langle \tau_{\alpha} \tau_{\beta} \rangle - \langle \tau_{\alpha} \tau_{\beta} \tau_{\delta} \rangle ) \equiv V.
\end{eqnarray*}
The cross terms in $ \nu $ and $ \varepsilon $ are
\begin{eqnarray*}
G_{\alpha \alpha} &=& \beta \sqrt{J_0 J_I} ( \langle \tau_{\alpha} \rangle \langle \sigma_{\alpha} \rangle -  \langle \tau_{\alpha} \sigma_{\alpha} \rangle ) \equiv G \\
G_{\alpha \beta} &=& \beta \sqrt{J_0 J_I} ( \langle \tau_{\alpha} \rangle \langle \sigma_{\beta} \rangle -  \langle \tau_{\alpha} \sigma_{\beta} \rangle ) \equiv H.
\end{eqnarray*}
and finally the cross terms in $ \rho $ and $ \eta $ are
\begin{eqnarray*}
G_{(\alpha \beta)(\alpha \beta)} &=& \beta^2 J J' ( \langle \tau_{\alpha} \tau_{\beta} \rangle \langle \sigma_{\alpha} \sigma_{\beta} \rangle - \langle \tau_{\alpha} \tau_{\beta}  \sigma_{\alpha} \sigma_{\beta} \rangle ) \equiv X \\
G_{(\alpha \beta)(\alpha \delta)} &=&  \beta^2 J J' ( \langle \tau_{\alpha} \tau_{\beta} \rangle \langle \sigma_{\alpha} \sigma_{\delta} \rangle  - \langle \tau_{\alpha} \tau_{\beta}  \sigma_{\alpha} \sigma_{\delta} \rangle ) \equiv Y \\
G_{(\alpha \beta)(\delta \gamma)} &=&   \beta^2 J J' ( \langle \tau_{\alpha} \tau_{\beta} \rangle \langle \sigma_{\delta} \sigma_{\gamma} \rangle  - \langle \tau_{\alpha} \tau_{\beta}  \sigma_{\delta} \sigma_{\gamma} \rangle )  \equiv Z.
\end{eqnarray*}
Due to the symmetry of the matrix under the permutation of the replica indices, one can obtain a complete set of eigenvectors for general $n$. There are three types of eigenvectors: the first one is symmetric under the interchange of indices, resulting in four eigenvalues. The second set of eigenvectors is symmetric under interchange of all but one of the indices and gives four eigenvalues which degenerate into the first eigenvalues for $ n = 0 $. We find out that these eigenvalues are always positive. Finally, the last set of eigenvectors that is symmetric under the interchange of all but two of the indices provides the following eigenvalues
\begin{eqnarray}\label{eq20}
\lambda &=& \frac12 \Bigg( \Big(P - 2Q + R + L - 2M + N\Big)  \nonumber \\
& & \pm \Big( (P - 2Q + R + L - 2M + N)^2 \nonumber \\
& &- 4 \big[(P - 2Q + R)(L - 2M + N) \nonumber \\
& &- (X- 2Y + Z)^2 \big] \Big)^{\frac12} \Bigg)
\end{eqnarray}
The condition that $ \lambda $ must be positive leads to the inequality in Eq. \eqref{eq9} for $ J'=0 $.

\bibliography{ref_Izadi1}
\bibliographystyle{apsrev}

\end{document}